**Structure and Charge transfer Mechanism in $Y_{1-x}Ca_xBa_2Cu_3O_{7-\delta}$ through direct doping**

Shiva Kumar Singh,[1,2] M. Husain,[2] and V.P.S Awana[1,*]

[1]Quantum Phenomena and Application Division, National Physical Laboratory (*CSIR*), New Delhi-110012, India
[2]Department of Physics, Jamia Millia Islamia, New Delhi-110025, India

**Abstract**

Here we study the effect of Ca doping on charge transfer mechanism of polycrystalline $YBa_2Cu_3O_{7-\delta}$ compound. The samples of composition $Y_{1-x}Ca_xBa_2Cu_3O_{7-\delta}$ ($x$ = 0.00, 0.05, 0.10, 0.20 and 0.30) are synthesized through standard solid state reaction route. Carrier doping is controlled by annealing of samples in oxygen and subsequently in reducing atmosphere. Samples are investigated using resistivity, *dc* magnetization (*M-T*) and magnetization with field (*M-H*) measurements. With increase of Ca the transition temperature ($T_c$) decreases in oxygenated samples, whereas the same increases in reduced samples. Further reduction of samples at higher temperatures (> 600 $^0C$) though results in non-superconducting nature up to Ca concentration of $x = 0.20$, the x = 0.30 sample is superconducting below 30K. This provides a remarkably simple and effective way to study the relationship between structure, superconductivity, and associated electronic properties. Variations in $Cu_1$-$O_4$, $Cu_2$-$O_4$ and $Cu_2$-$O_2$ bond lengths with oxygen content, is seen through the structural refinement of *XRD* pattern. The effective coordination of $Cu_2$ atom with oxygen changes with the change in these bond lengths and hence the holes in the $CuO_2$ planes. The charge transfer mechanism from $CuO_x$ chains to $CuO_2$ planes and thus effective hole doping is discussed in context of observed results.

PACS: 74.25.Wx, 74.25.fc, 74.25.Sv, 74.25.Op

Key Words: Superconductivity, Hole doping, Structural refinement, Bond length.

*Corresponding Author's awana@mail.nplindia.ernet.in*
Tel.: +91 11 45609357; Fax; +91 11 45609310
[†] *Web page- www.freewebs.com/vpsawana/*

## Introduction

The induction of superconductivity in $YBa_2Cu_3O_{7-\delta}$ (Y-123) cuprate can be achieved through the doping of holes in $CuO_2$ planes. In Y-123 structure $BaO/CuO_2/Y/CuO_2/BaO$ slabs are interconnected through $CuO_x$ having variable composition of oxygen [1-3]. The oxygen sites in the $CuO_2$ planes are identified as $O_2$ and $O_3$. The oxygen site in the BaO plane is termed as $O_4$. The $CuO_x$ chains oxygen sites are named as $O_1$ and $O_5$. The Y plane is devoid of oxygen. In tetragonal Y-123 structure, $O_2$, $O_3$ and $O_1$, $O_5$ are indistinguishable. Any contravene in integral $CuO_2$ stacks, affects superconductivity drastically [2-3]. The doping of holes can be done either by doping of electron in $CuO_x$ chains through the increase of oxygen content or by doping of electrons at Y site. In order to investigate the mechanism of superconductivity in $CuO_2$ planes divalent substitution at $Y^{3+}$ site, which is also mentioned as direct doping, may give some insight. It is expected that by such substitution, the charge distribution in the structure and consequently the concentration of charge carriers could be varied mostly at wish [4]. Ca is appropriate element for doping of electron at Y site as in divalent state it has ionic radii comparable to $Y^{3+}$. This leads to many interesting physical property changes in YBCO such as critical temperature ($T_c$), normal state conductivity ($\rho_n$), thermoelectric power and lattice parameters etc. $Ca^{2+}$ substitution for $Y^{3+}$ increases the overall hole concentration [5, 6]. The same substitution also leads to enhancement of $T_c$ in quenched samples of $PbSr_2(Y_{1-x}Ca_x)Cu_2O_{7+\delta}$ system [7-8]. However, partial substitution also occurs at Barium site in higher calcium concentrations [9]. In several studies [4, 10-11], it is found that decrease in the $T_c$, of fully oxygenated samples and the $T_c$ depression in such a situation can be ascribed to the overdoping effect [9, 12-13]. Excess holes introduced by $Ca^{2+}$ substitution also leads to increase in $J_c$ [14]. Fisher *et al*. [4] calculated the hall number at different oxygen contents and concluded that the hole doping by calcium is not uniform and depend on oxygen content of the system. Hole doping is very low for highly oxygenated samples, while it is higher for oxygen deficient ones. It is seen that in low δ regime, the compensating effect of calcium for hole filling is marginal [5, 6]. Calcium also alters the carrier density and the amount of trapped charge at the grain boundary and thus changing the form of the potential barrier [15]. Atomic-scale electron microscopy and transport measurements have been done on Ca-doped Y-123 bicrystals and improved flux pinning is reported due to Ca segregation to the grain boundary [16].

Along with these, Ca substitution also causes oxygen deficiency in doped samples. However, increase in $T_c$ is observed in Ca doped samples when it is made more oxygen deficient [11, 17-19]. Through neutron diffraction study it is observed that in oxygen deficient Y-123 that $Cu_1$-$O_4$ and $Cu_2$-$O_4$ bonds play important role in charge transfer and occurrence of $T_c$ [20]. It has been proposed that the Cu-O bonds which link the $CuO_x$ one-dimensional chains to the $CuO_2$ two-dimensional planes [the $Cu_1$-$O_4$ and $Cu_2$-$O_4$ bonds] offer a sensitive probe of the amount of charge transfer from the chains to the planes [21]. In neutron diffraction study of Ca doped Y-123 compounds it is found that overall number of holes increases with doping [17]. It was found that $Cu_2$-$O_2$ bond length has a deciding role in reduced samples.

In this paper we investigate the role of Ca, which leads to reduction of $T_c$ for fully oxygenated samples and further enhances the same for oxygen deficient samples. The manner in which the electronic properties changes with carrier doping is crucial for understanding the underlying mechanism of high-temperature superconductivity. Hole concentration changes with $c$-parameter and it has been evaluated using the unit cell length in the $c$ direction in the $CuO_2$ planes of Y-123 with a simple estimation [22]. However, the situation is different here with the presence of Ca and hence doping mechanism of holes is a bit complicated. The mechanism is discussed with help of observed results of structural changes being obtained through refinement.

The calculated bond lengths of $Cu_1$-$O_4$, $Cu_2$-$O_4$ and $Cu_2$-$O_2$ vary with oxygen content in the system. The effective coordination of $Cu_2$ atom with oxygen changes with the change in theses bond lengths and thus the carriers in the $CuO_2$ planes.

**Experimental**

The samples are synthesized in air by solid-state reaction route. The stoichiometric mixture of $CaCO_3$, $BaCO_3$, $Y_2O_3$, and $CuO$ are ground thoroughly, calcined at 880ºC for 12h and then pre-sintered at 910ºC and 925ºC for 20h with intermediate grindings. Finally, the powders are palletized and sintered at 925ºC for 20h in air. One set of as synthesized samples are annealed in oxygen atmosphere at $650^0$C and then at $450^0$C for 12 h each and then slowly cooled to room temperature. Two other set of the samples are created by annealing samples in reducing ($N_2$) atmosphere one at $450^0$C for 15 h and other at $600^0$C for 15 h. The phase formation is checked for each sample with powder diffractometer, Rigaku (Cu-Kα radiation) at room temperature. The phase purity analysis and lattice parameter refining are performed by Rietveld refinement programme (Fullprof version). The resistivity measurements of all samples are measured with standard four-probe method using APD cryogenics Closed Cycle Refrigerator. The magnetization measurements are carried out applying a field magnitude up to 1T using Physical Properties Measurement system- Quantum Designed PPMS-14T.

**Results and discussion**

All the samples are crystallized in nearly single phase which is confirmed from the Rietveld refinement of powder *XRD* patterns. However, some impurity phases are also observed in higher concentrations of Ca ($x = 0.20$ and $0.30$). A structural change has been observed with both increasing $x$ and with annealing condition [Tables1, 2 and 3]. In case of oxygen annealed samples all the compositions crystallized in orthorhombic *p/mmm* space group [23]. On the other hand for the samples reduced at $450^0$C although the Rietveld fitting is done [see Fig. 1(a)] with orthorhombic *p/mmm* space group, it infers that crystallization of pristine sample tends towards tetragonal *P4/mmm* space group for $x = 0.30$ composition [Table 3]. The set of samples reduced at $600^0$C are fitted in tetragonal *P4/mmm* space group [Fig. 1(b)]. An increase in *a*-parameter and decrease in *b*-parameter with increasing $x$ is seen, which can be explained on the basis of reduction of overall oxygen with increase in Ca content [4]. The O vacancies thus created lead towards *P4/mmm* tetragonal structure and thus both *a*- and *b*-parameters come closer. The increase in *c*-parameter with increase in $x$ is due to larger ionic radii of $Ca^{2+}$ (1.12 Å CN VIII) than $Y^{3+}$ (1.019 Å CN VIII). This also confirms successful substitution by $Ca^{2+}$ at $Y^{3+}$site. Slight decrease/no change in lattice parameter has also been observed earlier [10] and it was argued that Ca goes with six coordination of oxygen rather than that of eight coordination of Y with oxygen depending upon overall oxygen content of the system. It seems O vacancies induced in Ca doped samples are playing a crucial role. The coordination of Ca with oxygen might be effected with insertion of these vacancies. In case of reduced samples the changes in lattice parameters can be explained as: The oxygen deficiency due to reduction of samples is responsible for increase in *a*- and decrease in *b*-parameter, in comparison to oxygenated samples. Also, increase in *c*-parameter is observed in reduced samples and it increases with reduction temperature [see table]. The increase of *c*-parameter in oxygen deficient samples is in agreement with ref. [17, 20, 24]. The reason that leads to increase in *c*-parameter will be discussed later.

The resistivity plot of all the samples shows that the $T_c$ value decreases with increasing Ca content [Fig. 2]. The simultaneous variation of lattice parameters and $T_c$ with $x$ is also a confirmation of substitution by Ca at Y site. The samples up to $x = 0.2$ show sharp transition but $x = 0.3$ concentration exhibits relatively broader transition with two steps. This may be due to partial decomposition taking place in the sample with $x = 0.3$ [2, 23]. Ca addition also induces disorder in the superconducting $CuO_2$ planes [13]. Up to a critical oxygen content ($x_c$) disorder of the oxygen remains ineffective but it creates mobile hole traps when $x > x_c$ [11]. The value of $x_c$ gets lower with increase of Ca content. Therefore reduction in $T_c$ with increasing Ca content is the collective effect of over-doping along with the oxygen disorder. On the other hand $Cu_2$-$O_2$-$Cu_2$ bond angle increases with the increase of carriers in $CuO_2$ planes [25] which causes flattening of $CuO_2$ planes. This flattening, results in decrease of internal crystallographic pressure of the system which also leads to decrease in $T_c$ [26-27]. Inset of Fig. 2 shows that magnetization measurements (*M-T*) of the samples annealed in oxygen, in both Field Cooled (*FC*) and Zero Field Cooled (*ZFC*) conditions. It can be seen that $T_c$ decrease is mimicking the resistivity behaviour. The saturated diamagnetic magnetization, which is measure of diamagnetic shielding current increases systematically up to $x = 0.20$. Some of us earlier found that a pinning like behavior is developed with Ca doping and therefore an enhancement of the critical current density ($j_c$) [23]. The increase in separation of *FC* and *ZFC* signals with Ca content in doped samples indicates a strong flux trapping process. The separation of *FC* and *ZFC* signals increases monotonically until $x = 0.20$ sample and nearly saturates for $x = 0.30$.

Fig. 3 (a) shows the *M-T* of the samples annealed in nitrogen atmosphere at $450^0C$ in *FC* and *ZFC* conditions. The situation is reversed now than that of in oxygenated samples. The enhancement in $T_c$ can be seen with increase of Ca content. This is in agreement with earlier studies [11, 17-19]]. The same is also inferred from the *M-H* loops of the samples [Fig. 3 (b)]. What is the mechanism that leads to this remarkable change in $T_c$ is need to be emphasized.

Actually the doping mechanism of holes in $CuO_2$ planes is changed now [to be discussed]. It is considered that hole transfer from $CuO_x$ chains to $CuO_2$ planes is done through the apex oxygen atom O4 [20-21]. Thus $Cu_1$-$O_4$ and $Cu_2$-$O_4$ bond lengths are needed to be focused. The $Cu_1$-$O_4$ and $Cu_2$-$O_4$ bond lengths determined through the Rietveld refinement of the *XRD* pattern of the samples are given in the table1, 2 and 3. In case of reduced samples decrease in $Cu_1$-$O_4$ bond length along with increase in $Cu_2$-$O_4$ bond length is observed [see table2 and 3]. Similar pattern is also observed in earlier studies [17, 20-21]. The increase in $Cu_2$-$O_4$ bond length is larger than decrease in $Cu_1$-$O_4$ bond length. This may be responsible for increase in *c*-parameter of reduced samples. Fig. 4 shows the variation of *dc* magnetization with temperature in *FC* and *ZFC* conditions of the $Y_{1-x}Ca_xBa_2Cu_3O_{7-\delta}$ ($x = 0.00$ and 0.30) samples. Inset shows *M-H* curves of the same samples at 5 K. It shows that further reduction almost kills superconductivity or very less volume fraction up to $x = 0.20$ and is retained at round 30 K in $x = 0.30$ composition.

The mechanism of hole doping can be understood in terms of effective oxygen coordination of $Cu_2$ atoms. It seems in case of oxygen annealed samples apex oxygen $O_4$ is playing important role, on the other hand, in case of reduced samples the in plane ($CuO_2$ plane) oxygen $O_2$ is getting prominent role along with the $O_4$, for hole doping. It can be seen that in oxygenated samples there is increase in $Cu_1$-$O_4$ bond length whereas a decrease in $Cu_2$-$O_4$ bond length with *x*. Decrease in $Cu_2$-$O_2$ bond length with *x* can also be seen. This is a clear evidence of overdoping in $CuO_2$ planes. This implies that effective coordination of $Cu_2$ atom with oxygen is increasing and hence injection of extra holes in $CuO_2$ planes. In case of reduced samples the $Cu_1$-$O_4$ bond length is smaller than that of oxygenated samples and also decreasing with *x*. At the same time $Cu_2$-$O_4$ bond length is larger than that of oxygenated samples and increasing with *x*. Though the $Cu_2$-$O_4$ bond length is increasing with *x*, the $Cu_2$-$O_2$ bond length is decreased [see tables]. This is in agreement with the earlier findings of neutron diffraction study of reduced sample by some of us [17]. Thus the effective coordination of $Cu_2$ atom with oxygen is still increasing and resulting in optimum doping of holes. This way one gets increased superconductivity in moderately reduced Ca doped samples and induction of superconductivity of up to 30K in highly reduced samples. Our results provide clear evidence through structural details that mobile hole carriers can be directly induced by $Y^{3+}/Ca^{2+}$ substitution in non superconducting highly reduced $YBa_2Cu_3O_7$ (Y-123).

**Conclusion**

Direct doping of carriers by $Y^{3+}/Ca^{2+}$ substitution in Y-123 system at varying oxygen concentration is studied in detail. Both decrease/increase and appearance of superconductivity is observed respectively in optimum oxygenated, moderately and highly reduced samples. Detailed structural analysis show that the calculated bond lengths of $Cu_1$-$O_4$, $Cu_2$-$O_4$ and $Cu_2$-$O_2$ vary with oxygen content in the system. The effective coordination of $Cu_2$ atom changes with the change in bond lengths and thus the carriers in the $CuO_2$ planes. The charge transfer mechanism from $CuO_x$ chains to $CuO_2$ planes and hole doping in $CuO_2$ planes is discussed in unique combination of both change in over all oxygen content and the direct doping by $Y^{3+}$ site $Ca^{2+}$ substitution.

**Acknowledgements**

The authors would like to thank DNPL Prof. R. C. Budhani for his constant support and encouragement. One of the authors Shiva Kumar would like to acknowledge *CSIR*-India for providing fellowships.

# References

[1] M. Karppinen, V.P.S. Awana, Y. Morita, H. Yamauchi, Physica B, 312-313, 62 (2003).

[2] P.R. Slater, C. Greaves, Physica C 180, 299 (1991).

[3] Shiva Kumar, Anjana Dogra, M. Husain, H. Kishan and V.P.S. Awana, J. Alloys and compd. 352, 493 (2010).

[4] B. Fisher, J. Genossar, C. G. Cuper, L. Patlagan, G. M. Riesner and A. Kniizhnik, Phys. Rev. B. 47, 6054 (1992).

[5] C. Greaves and P.R. Slater, Supercond. Sci. Technol. 2, 5 (1989).

[6] R. G. Buckley, J. L Tallon, D. M. Pooke and M.R. Presland, Physica C 165, 391 (1990).

[7] Shiva Kumar, Monika Mudgel, M. Husain, H. Kishan and V.P.S. Awana, Physica C 470 S205 (2010).

[8] T. Maeda K. Sakuyama, S. Koriyama, H. Yamauchi, and S. Tanaka, Phys. Rev. B 43 7866 (1991)

[9] R. G. Buckley, D. M. Pook, J. L. Tellon, M. R. Persland, N. E. Flower, M. P. Staines, H. L. Johnson, M. Meylan, G. V. M. Williams, and M. Wowden, Physica C 174, 383 (1993).

[10] V. P. S. Awana and A. V. Narlikar, Phys. Rev. 8 49, 6385 (1994).

[11] A. Manthiram and J. B. Goodenough, Physica C 159, 760 (1989).

[12] Kiyotaka Matsura, Takihiro Wada, Yujii Yakeashi, S. Tajiima, and H. Yamauchi, Phys. Rev. B 46, 11923 (1992).

[13] R. Nagarajan, Vikram Pavate, and C. N. R. Rao, Solid State Commun. 84, 183 (1993).

[14] Kucera, J. T. & Bravman, J. C. Phys. Rev. B 51, 8582 (1995).

[15] J. Mannhart and H. Hilgenkamp, Mater. Sci. Eng. B 56, 77 (1998).

[16] X. Song, G. Daniels, D. Matt Feldmann, Alex Gurevich and D. Larbalestier Nat. mats. 4 (2005) 470.

[17] V. P. S. Awana, S. K. Malik, and W. B. Yelon, Physica C 262, 272 (1996).

[18] V. P. S. Awana, A. Tulapurkar, and S. K. Malik, Phys. Rev. B 50, 594 (1994).

[19] E. M. McCarron III, M. K. Crawford, and J. B. Parise, J. Solid State Chem. 78, 192 (1989).

[20] J. D. Jorgensen, B.W. Veal, A. P. Paulikas, L. J. Nowicki, G. W. Crabtree, H. Claus, and W. K. Kwok Phys. Rev. B 41, 1863 (1990).

[21] I. F. David, T. A. Harrison, J. M. F. Gunn, O. Moze, A. K. Soper, P. Day, J. D. Jorgensen, M. A. Beno, D. W. Capone, D. G. Hinks, I. K. Schuller, L. Soderholm, C. U. Segre, K. Zhang, and J. D. Grace, Nature 327, 310 (1987).

[22] R. Liang, D. A. Bonn, W. N Hardy, Phys. Rev. B 73, 180505 (2006).

[23] N.P Liyanawaduge, Anuj kumar, Shiva Kumar, B.S.B Karunarathne V.P.S Awana, J. Sup. Nov. Magn. 25, 31 (2012).

[24] R.J. Cava, B. Batlogg, C.H Chen, E. A. Rietman, S.M Zuhurak and D. Werder, Nature, 329, 423 (1987).

[25] S. R. Ovshynsky, Chem. Phy. Lett. 195, 455(1992).

[26] H. Shakeripour, M. Akhawan, Supercond. Sci. Technol. 14, 213(2001).

[27] Y. K Wu, J. R, Asbarn, C. J. Tong, P. H. Hor, R.L. Meng, L. Gao, Z. J. Hung, Y.Q. Wang and C.W Chu, Phys. Rev. Lett 58, 908 (1987).

**Table Caption**

**Table 1**: Rietveld Refined lattice parameters, Cu1-O4, Cu2-O4 and Cu2-O2 bond lengths of oxygen annealed $Y_{1-x}Ca_xBa_2Cu_3O_{7-\delta}$ ($x$ = 0.00, 0.10, 0.20 and 0.30) samples.

**Table 2**: Rietveld Refined lattice parameters, Cu1-O4, Cu2-O4 and Cu2-O2 bond lengths of $450^0$ C nitrogen annealed $Y_{1-x}Ca_xBa_2Cu_3O_{7-\delta}$ ($x$ = 0.00, 0.10, 0.20 and 0.30) samples.

**Table 3**: Rietveld Refined lattice parameters, Cu1-O4, Cu2-O4 and Cu2-O2 bond lengths of $600^0$ C nitrogen annealed $Y_{1-x}Ca_xBa_2Cu_3O_{7-\delta}$ ($x$ = 0.00, 0.10, 0.20 and 0.30) samples.

**Figure Caption**

Fig. 1(a): Rietveld fitted *XRD* pattern of $450^0$ C nitrogen annealed $Y_{1-x}Ca_xBa_2Cu_3O_{7-\delta}$ ($x$ = 0.00, 0.10, and 0.30) samples with space group *Pmmm*.

Fig. 1(b): Rietveld fitted *XRD* pattern of $600^0$ C nitrogen annealed $Y_{1-x}Ca_xBa_2Cu_3O_{7-\delta}$ ($x$ = 0.00, 0.10, and 0.20) samples with space group *P4/mmm*.

Fig. 2: Resistivity behavior with Temperature [$\rho(T)$] of $Y_{1-x}Ca_xBa_2Cu_3O_{7-\delta}$ ($x$ = 0.00, 0.10, 0.20 and 0.30) of oxygen annealed samples. Inset shows variation of *dc* volume susceptibility with temperature in *FC* and *ZFC* conditions of the same samples.

Fig. 3(a): Variation of *dc* magnetization [*M-T*] with temperature in *FC* and *ZFC* conditions of the $Y_{1-x}Ca_xBa_2Cu_3O_{7-\delta}$ ($x$ = 0.00, 0.05, 0.10, 0.20 and 0.30) samples annealed at $450^0$ C in nitrogen atmosphere.

Fig. 3(b): *M-H* curves of the $Y_{1-x}Ca_xBa_2Cu_3O_{7-\delta}$ ($x$ = 0.00, 0.05, 0.10, 0.20 and 0.30) samples at 10 K up to magnetic field 1T, annealed at $450^0$ C in nitrogen atmosphere.

Fig. 4: Variation of *dc* magnetization [*M-T*] with temperature in *FC* and *ZFC* conditions of the $Y_{1-x}Ca_xBa_2Cu_3O_{7-\delta}$ ($x$ = 0.00 and 0.30) samples annealed at $600^0$ C in nitrogen atmosphere. Inset shows *M-H* curves of the $Y_{1-x}Ca_xBa_2Cu_3O_{7-\delta}$ ($x$ = 0.00 and 0.30) samples at 5 K.

**Table 1**

| $Y_{1-x}Ca_xBa_2Cu_3O_{7-\delta}$ $O_2$ ann. | a(Å) | b(Å) | c(Å) | $R_p$ | $R_{wp}$ | $\chi^2$ | Cu1-O4 (Å) | Cu2-O4 (Å) | Cu2-O2 (Å) |
|---|---|---|---|---|---|---|---|---|---|
| $x = 0.0$ | 3.822(1) | 3.887(2) | 11.686(4) | 5.08 | 6.57 | 3.40 | 2.003(3) | 2.247(4) | 1.977(3) |
| $x = 0.1$ | 3.826(7) | 3.881(4) | 11.694(5) | 4.91 | 6.28 | 2.89 | 2.011(2) | 2.216(3) | 1.973(2) |
| $x = 0.2$ | 3.831(4) | 3.877(5) | 11.703(4) | 4.80 | 6.19 | 2.63 | 2.030(4) | 2.160(4) | 1.971(2) |
| $x = 0.3$ | 3.836(2) | 3.873(3) | 11.711(6) | 5.10 | 6.60 | 3.65 | 2.033(2) | 2.136(2) | 1.970(4) |

**Table 2**

| $Y_{1-x}Ca_xBa_2Cu_3O_{7-\delta}$ 450 C $N_2$ ann. | a(Å) | b(Å) | c(Å) | $R_p$ | $R_{wp}$ | $\chi^2$ | Cu1-O4 (Å) | Cu2-O4 (Å) | Cu2-O2 (Å) |
|---|---|---|---|---|---|---|---|---|---|
| $x = 0.0$ | 3.842(3) | 3.872(3) | 11.752(4) | 5.30 | 6.94 | 3.85 | 1.945(4) | 2.263(4) | 1.957(6) |
| $x = 0.1$ | 3.857(1) | 3.860(6) | 11.768(5) | 5.16 | 6.60 | 2.95 | 1.932(5) | 2.275(1) | 1.931(3) |
| $x = 0.2$ | 3.856(3) | 3.861(7) | 11.781(6) | 4.75 | 6.14 | 2.83 | 1.918(2) | 2.290(5) | 1.924(1) |
| $x = 0.3$ | 3.852(6) | 3.860(2) | 11.793(6) | 5.29 | 6.89 | 3.34 | 1.905(1) | 2.334(6) | 1.924(4) |

**Table 3**

| $Y_{1-x}Ca_xBa_2Cu_3O_{7-\delta}$ 600 C $N_2$ ann. | a(Å) | b(Å) | c(Å) | $R_p$ | $R_{wp}$ | $\chi^2$ | Cu1-O4 (Å) | Cu2-O4 (Å) | Cu2-O2 (Å) |
|---|---|---|---|---|---|---|---|---|---|
| $x = 0.0$ | 3.861(2) | 3.861(2) | 11.810(6) | 5.25 | 6.62 | 3.34 | 1.889(2) | 2.404(3) | 1.943(2) |
| $x = 0.1$ | 3.858(1) | 3.858(1) | 11.820(7) | 4.53 | 5.67 | 2.46 | 1.872(2) | 2.399(5) | 1.943(5) |
| $x = 0.2$ | 3.859(6) | 3.859(6) | 11.824(9) | 4.99 | 6.35 | 2.93 | 1.863(6) | 2.405(3) | 1.939(2) |
| $x = 0.3$ | 3.854(7) | 3.854(7) | 11.828(7) | 5.69 | 6.46 | 3.95 | 1.868(1) | 2.401(6) | 1.938(3) |

Fig. 1(a)

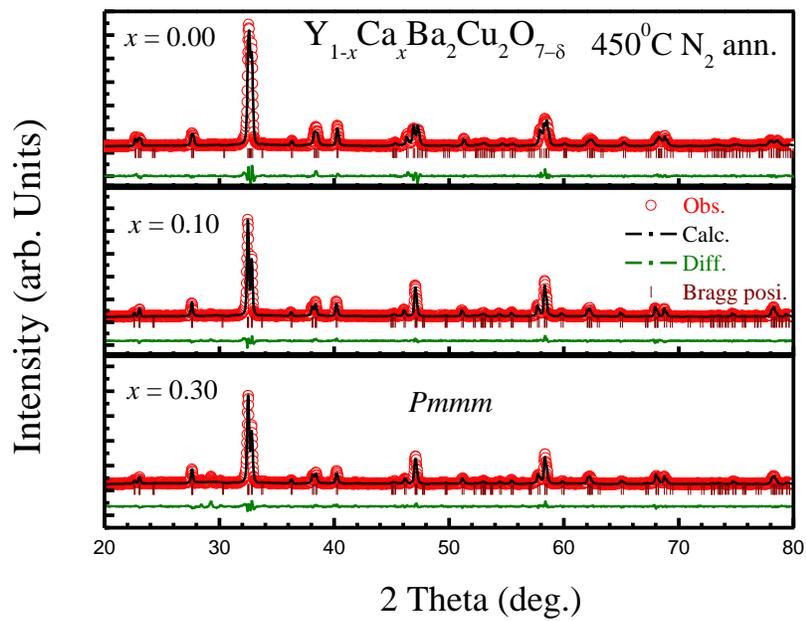

Fig. 1(b)

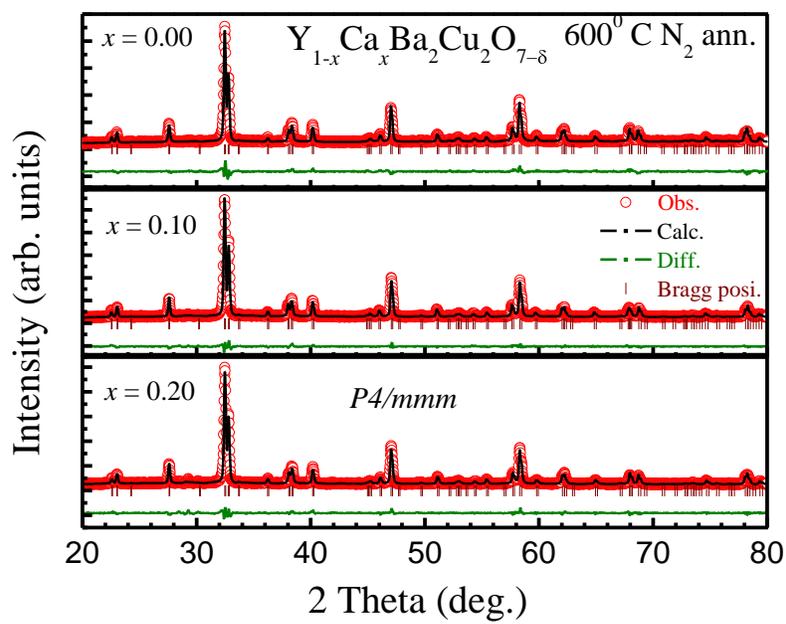

Fig. 2

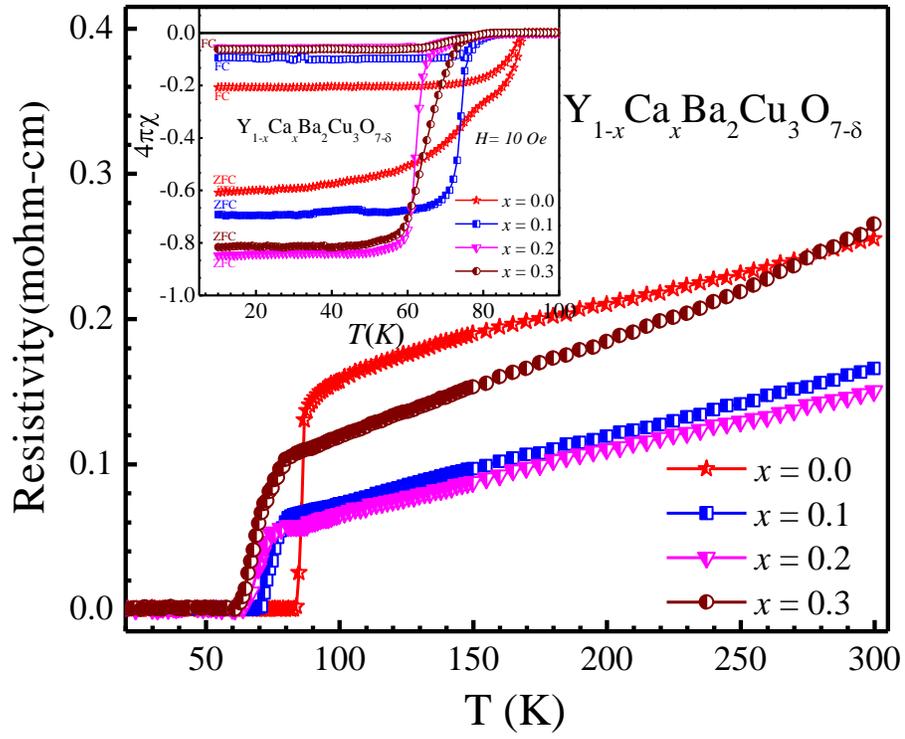

Fig. 3(a)

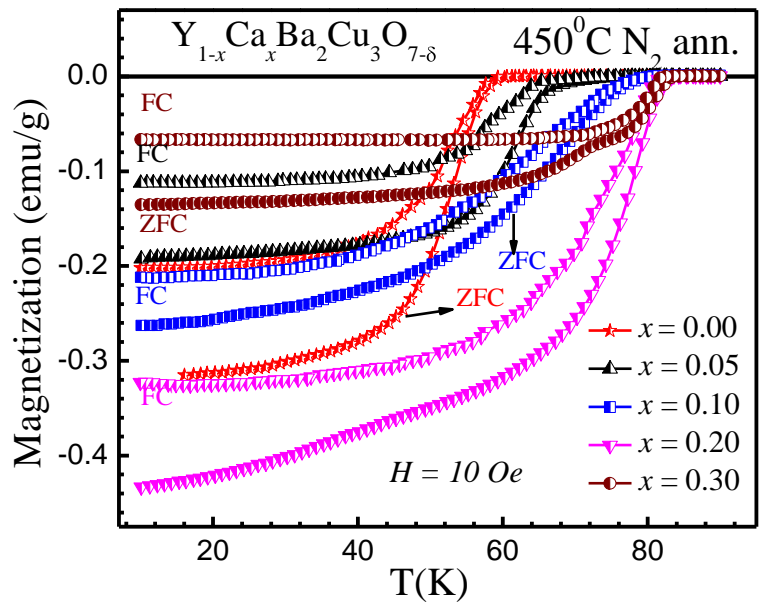

Fig. 3(b)

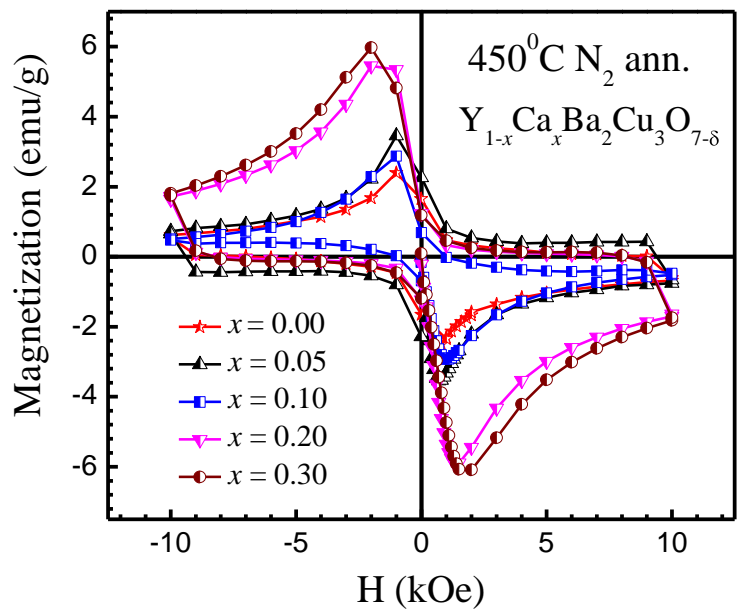

Fig. 4

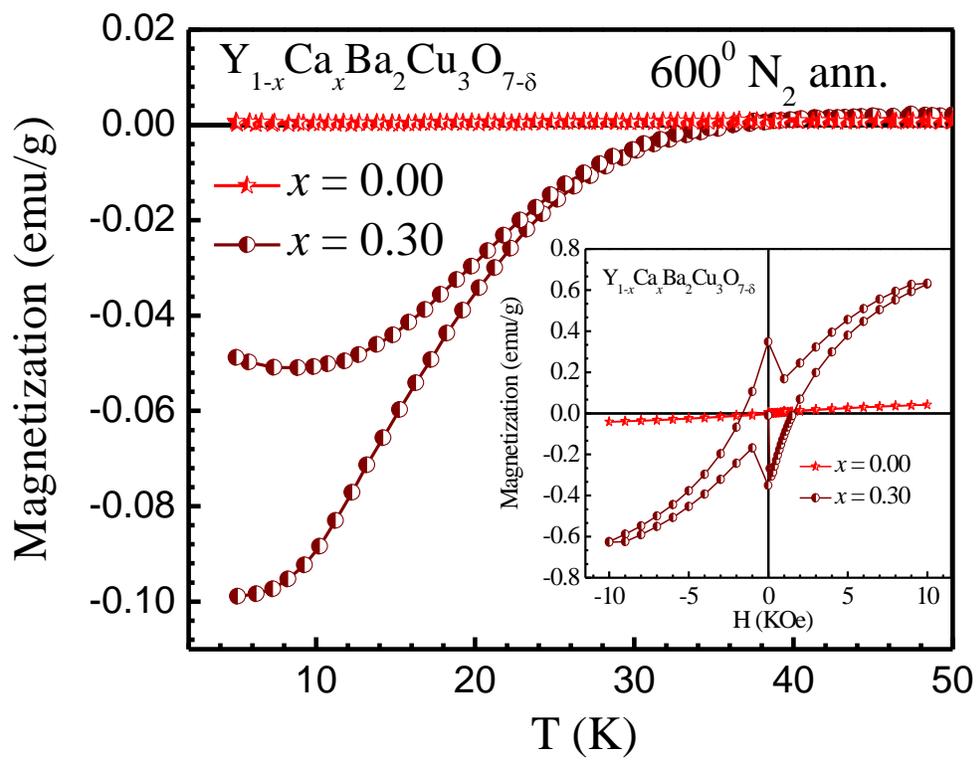